\documentstyle[11pt,aaspp4]{article}

\lefthead{Boroson et al.}
\righthead{HUT Observations of Her X-1}
\begin{document}
  
\title{Hopkins Ultraviolet Telescope Observations of Her X-1} 

\author{Bram Boroson}
\affil{Center for Astrophysics, 60 Garden Street, Cambridge, MA 02138;
bboroson@cfa.harvard.edu}

\author{William P. Blair, Arthur F. Davidsen}
\affil{Department of Physics and Astronomy, The Johns Hopkins University,
Charles \&\ 34th Streets, Baltimore, MD 21218; wpb@pha.jhu.edu,
afd@pha.jhu.edu}

\author{S.D. Vrtilek, John Raymond}
\affil{Center for Astrophysics, 60 Garden Street, Cambridge, MA 02138;
svrtilek@cfa.harvard.edu, 
jraymond@cfa.harvard.edu}

\author{Knox S. Long}
\affil{Space Telescope Science Institue, 3700 San Martin Drive, 
Baltimore, MD 21218; long@stsci.edu}

\and

\author{Richard McCray}
\affil{JILA, Campus Box 440, University of Colorado, Boulder, CO 80309;
dick@jila.colorado.edu}
 
\newcommand{\Msun}{\mbox{ M}_{\odot}}  
\newcommand{\lae}{\mathrel{<\kern-1.0em\lower0.9ex\hbox{$\sim$}}}
\newcommand{\gae}{\mathrel{>\kern-1.0em\lower0.9ex\hbox{$\sim$}}}
\newcommand{\kms}{km s$^{-1}$}
\newcommand{\NV}{N\,V\,$\lambda\lambda\,1238.8,\,1242.8$}
\newcommand{\CIV}{C\,IV\,$\lambda\lambda\,1548.2,\,1550.8$}
\newcommand{\SiIV}{Si\,IV\,$\lambda\lambda\,1393.7,\,1402.8$}
\newcommand{\OVI}{O\,VI\,$\lambda\lambda\,1031.9,\,1037.6$}

\begin{abstract} 

We have obtained a far-ultraviolet spectrum of the X-ray binary
Hercules~X-1/HZ~Herculis using the Hopkins Ultraviolet Telescope aboard
the Astro-1 space shuttle mission in 1990 December.  This is the first
spectrum of Her X-1 that extends down to the Lyman limit at 912~\AA.  We
observed emission lines of \OVI, \NV, and \CIV, and the far UV continuum
extending to the Lyman limit.  We examine the conditions of the emitting
gas through line strengths, line ratios, and doublet ratios.  The UV flux
is lower by about a factor of~2 than expected at the orbital phase of the
observation.  We model the UV continuum with a simple power-law and with a
detailed model of an X-ray-illuminated accretion disk and companion star. 
The power-law provides a superior fit, as the detailed model predicts too
little flux below 1200\AA.  We note, however, that there are
uncertainties in the interstellar reddening, in the background 
airglow spectrum, and in the long-term phase
of the accretion disk.  We have searched the data for UV~line and
continuum pulsations near the neutron star spin period but found none at a
detectable level. 

\end{abstract}  
 
  
\section{Introduction}
 
Hercules X-1/HZ Herculis is a low mass X-ray binary system consisting of a
$\approx1.3\Msun$ neutron star and a $\approx2.2\Msun$ A dwarf star locked
in orbit about each other with a period of 1.7~d (Deeter et al. 1991, and
references therein).  The X-ray source is eclipsed by the A star for
0.24~d of each period, indicating an orbital inclination near 90$^\circ$. 
The spectral type of the A star varies with orbital phase from late
B/early A to late A/early F such that the later spectral types correspond
to eclipse and the earlier spectral types correspond to 1/2 an orbit later
(when X-ray heating of the A star should be most apparent--see Milgrom \&
Salpeter 1975).  Mass transfer, which may be periodic, feeds an accretion
disk around the neutron star. Accretion onto the neutron star itself,
which has a magnetic field of $\sim4\times10^{12}$ Gauss, occurs on the
magnetic poles (Gruber et al. 1980; Voges et al. 1982). 
 
The X-rays from the system vary on a period near 34.85 d (\"Ogelman 
1987, Baykal et al. 1993), with an X-ray high state typically lasting
$\approx$11~d, and an X-ray low state (with a secondary maximum of 30\%\
the high state) lasting $\approx$24~d.  The time at which the X-ray high
state begins varies unpredictably 
from the mean ephemeris by more than
$\pm1$~d, an effect that has led to the generation of a number of
different models (e.g. Petterson 1975; Gerend \& Boynton 1976; Wolff \&
Kondo 1978; Kondo, Van Flandren, \& Wolff 1983; Meyer \& 
Meyer-Hofmeister 1984; Schandl \& Meyer 1994).  As the
1.7~day modulation of the optical light continues during the
X-ray low states (Bahcall \&\ Bahcall 1972), the 35-day cycle probably
results from a geometric effect, such as obscuration by a precessing 
accretion disk, and not an interruption of the accretion, which
would cause the X-ray heating of HZ~Her to cease.
In addition to the 35~d cycle, pulsed X-ray emission with a period
of 1.24s is also seen, which is generally interpreted as the rotation period 
of the neutron star (Tananbaum et al. 1972; Giacconi et al. 1973).  These
pulsations have also been detected optically (Davidsen et al. 1972;
Middleditch 1983), in the UV (Boroson et al. 1996) and at optical
and infrared wavelengths (Middleditch, Puetter, \& Pennypacker 1985). 
 
The ultraviolet spectrum above 1200~\AA\ has been extensively observed
with IUE (Dupree et al. 1978; Gursky et al. 1980; Howarth \& Wilson
1983a,b; Vrtilek \&\ Cheng 1996).  Both the line and continuum components 
of the spectrum vary with orbital phase.
The ultraviolet light from Her X-1 comes from at least two main 
regions within the system:  the heated face of the normal star and the 
accretion
disk.  According to the analysis of Howarth \& Wilson (1983a), roughly 
1/2 of the hard X-rays striking the A~star are reprocessed into UV and 
optical continuum emission.  Reprocessed X-rays are also responsible
for most of the UV light from the disk.  Using the Goddard High 
Resolution Spectrograph ({\it GHRS}) aboard the Hubble Space 
Telescope, Boroson et al. (1996) showed that the UV lines have
broad and narrow components which they interpreted in terms of 
emission from the disk and the star.
 
In this paper, we present a spectrum of Her X-1 that extends the
far-ultraviolet wavelength coverage to the Lyman limit at 912~\AA.
We have detected, for the first time, the \OVI\ emission lines, and the 
continuum in this wavelength region.
In  \S2\ we discuss the
observations and data reduction, and in \S3\ we discuss
our data in the context of other observations of Her~X-1, and we 
compare the data with models of the line and continuum emission.
 
\section{Observations}
 
The observations were carried out with the Hopkins Ultraviolet Telescope
(HUT) as part of the Astro-1 space shuttle mission in 1990 December.
HUT consists of a 0.9~m mirror that feeds a prime focus spectrograph
with a microchannel plate intensifier and reticon detector.  First order
sensitivity covers the region from 850~\AA\ to 1850~\AA\ at 
0.51~\AA~pix$^{-1}$, with $\approx$3\AA\ resolution.  The detector can be 
operated
in any of several modes, including a high time resolution mode where individual
photons can be time-tagged to a relative accuracy of 1~ms and an absolute
accuracy of 3~ms.
This was the mode of operation during the observation described here.
Further details of the spectrograph and telescope and their on-orbit 
performance and calibration can be found in Davidsen et al. (1992)
and Kruk et al. (1997).
 
The Her X-1 observation discussed here took place on 1990 December~6 at
4:22~UT.  According to the 1.7 day orbital ephemeris of Deeter et al. (1991),
this corresponds to $\phi_{\rm orb}=0.66$.  We do not know the X-ray state
of Her~X-1 at the time of our observation.
Using a period $P_{35}=34.9$~d and the observed X-ray turn-on nearest in 
time to our observation (JD~2448478.3, as given in Baykal et al. 1993), our 
observation took place at $\phi_{35}=0.93$, while X-ray turn-on occurs at
$\phi_{35}=0.0$.  This suggests that our observation took place in the 
X-ray off state; however, the 
times of the X-ray turn-ons are known to vary by $\sim$6 days (\"Ogelman
1987; Baykal et al. 1993), so that without simultaneous X-ray observations,
we can not be certain.
 
The HUT observation was made with an 18$\arcsec$ aperture placed
on the object, and guiding was accomplished using preplanned guide
star positions on the HUT TV guider.  The total integration was 1700~s.
Orbital viewing 
constraints required that we observe Her~X-1 entirely during the orbital
day of HUT, which resulted in substantial airglow contamination,
leading not only to 
significant line emission (including several EUV lines seen in second 
order) but also a pseudo-continuum from grating-scattered light
(chiefly from the brightest airglow lines at Ly$\alpha$ and 
O\,I$\lambda1302$).

The flux calibration is based on observations of the white dwarf star
G191-B2B compared with model stellar atmospheres and laboratory
calibrations as discussed by Davidsen et al. (1992) and Kruk et al. (1997).
The relative fluxes are accurate to within 5\%.

The count rate in the continuum from Her X-1 (in the 1400--1600\AA\ range, 
relatively free from airglow emission) was not uniform,
but peaked at 5.5 counts~s$^{-1}$ during a five-minute interval.
This count rate should be compared with an average
rate of 4.3 counts~s$^{-1}$ (1400--1600\AA) in the 1500~s interval 
that we used for our analysis.
Data from the HUT TV camera indicate sections at the beginning and end of the
observation where pointing stability was disrupted.
Post-flight analysis of video frames stored during the observation show
that even during the portion of the observation where the pointing
stability was good, the object was not well centered in the slit,
especially in the coordinate perpendicular to the dispersion direction.
Hence, even minor pointing errors could have affected the count
rate (and derived flux) because of vignetting by the edge of the slit.   
(The PSF of the telescope produced star images with FWHM$\sim5\arcsec$.)  
We thus suspect that the fluctuating count rate was
due to pointing errors, and that the maximum observed count rate is
probably representative of the true flux of the object.
This analysis implies that the average count rate was low by 
$\approx$27\%; 
the derived fluxes (and errors) were thus multiplied by 1.27 to
correct for this effect.  Since the HUT detector is 
photon-counting, it is straightforward to generate and propagate errors
along with the data as various operations are performed.
To the extent that flux was being lost even during stable pointing (due
to vignetting) this corrected flux could still be too low, but not by    
more than $\sim20$\%.  The corrected, flux-calibrated spectrum is shown as
the upper curve in Figure~1.

In order to subtract airglow contamination to first order, we have fit and
subtracted a blank day sky observation obtained with the same aperture
during Astro-1.  We used smooth fits to the known airglow lines in this 
spectrum instead of the observed airglow background to reduce the
contribution of counting noise.  An actual smoothing of the background
spectrum, instead of a fit to individual lines, would have increased the
widths of the strong airglow lines where they overlap emission lines from
Her~X-1.  We have continued to use the error array associated with the
counts in the original data.  The result of our fit to the background
spectrum is shown as the lower curve in Figure~1. 

The airglow-subtracted spectrum
representing the summed intrinsic data from the source is shown in
Figure~2.
Because the relative intensities of the fainter airglow lines
change with orbital latitude and look-angle with respect to the
earth limb, the subtraction of line emission
is not perfect.  Also, a variable complex of airglow emissions affects the
spectrum in the region near the Lyman limit (cf. Feldman et al. 1992).
However, our technique should be sufficient to provide
a reasonable representation of the intrinsic continuum longward of 930~\AA.
This is the spectrum that will be used for continuum fitting below.
 
We have identified emission from the \NV\ doublet and the \CIV\ doublet,
which have been seen with IUE (Boyle et al. 1986, Howarth \& Wilson 1983a,
1983b) and HST (Anderson et al. 1994; Boroson et al. 1996), and also the
O\,VI$\lambda\lambda$1031.9,1037.6 resonance doublet and some fainter
emission lines at a more marginal detection level.  This 
constitutes the first detection of O~VI emission from the Her~X-1
system, and provides a significant new diagnostic of photoionized gas
in this system.  

In order to provide quantitative estimates of the line strengths,
we have fit Gaussian profiles to the data.  Since some of the lines,
including \OVI, are affected by nearby airglow features, we first
used our template airglow spectrum to carry out a local subtraction
of the airglow in order to produce a flat continuum adjacent to 
each line.  Typically, this required us to scale the template
spectrum by of order~20\%\ from the value used above to
achieve a reasonable local fit near the lines. 
 
Each doublet was fit assuming an optically thick 1:1 ratio for the lines
and a fixed separation of the doublet components (although the central
wavelength of each feature was allowed to vary). Observations with the HST
show two line components; for a narrow component, the doublet ratios are
typically $\approx1.3:1$, while for the broad component, the doublet ratio
is more nearly $1:1$.  We found a superior fit (reduced $\chi^2=1.51$
instead of $\chi^2=1.73$ with 42 degrees of freedom) when we allowed a
$2:1$ doublet ratio for \OVI. Allowing the doublet ratio to be a free
parameter does not improve the fit ($\chi^2=1.52$ for a ratio of 1.65:1). 
Alternately, the \OVI\ lines could have a 1:1 ratio, but an interstellar
C\,II$\lambda1037.3$ line could absorb O\,VI$\lambda1037.6$. Allowing the
wavelength of the interstellar line to vary, but fixing its width as the
width of the O\,VI lines (the instrumental width), we found $\chi^2=1.57$
with 42 degrees of freedom. The $\chi^2$ value for all the O\,VI fits may
be artificially high, as the errors do not reflect the uncertainty in the
subtraction of the nearby strong airglow lines of H~Ly$\beta$,
O\,I$\lambda1026$, and O\,I$\lambda1040$. Figure~3 shows the region near
\OVI\ with the observed and modelled airglow lines, and Figure~4 shows the
best fits for all the lines. 

In addition to \OVI, \NV, and \CIV, there was marginal (3$\sigma$)
evidence for an O\,V line at 1371~\AA, at a strength relative to
N\,V and C\,IV similar to that seen with the HST.  He\,II$\lambda1640$ 
may also have been detected at a marginal level.  There is no 
evidence for Si\,IV$\lambda1400$, and while some features appear at 
$\approx974$\AA, they are not significant enough to identify as 
C\,III$\lambda977$.  Any possible C\,III$\lambda1176$ is lost
in the airglow emission that affects this region of the spectrum.

All of the lines intrinsic to Her X-1 are blueshifted.  After detailed
inspection of the video frames, we were unable to determine unequivocably
whether or not this shift is real.  Based on analysis of two guide stars,
we infer an offset within the aperture in the dispersion direction that is
consistent with a blueshift of 0.0 -- 1.9~\AA\ (0.0 -- 5.6$\arcsec$
miscentering).  
The discrepant results from the different guide stars are
due to a combination of distortions within the TV camera and measurement
error.  If the blueshift caused by the miscentering is near 1.9~\AA, then 
the apparent blueshifts of the lines are not real, and it becomes more 
likely that interstellar C\,II$\lambda1037.3$ can absorb O\,VI$\lambda1037.6$.

We searched for pulsations in the UV lines from Her X-1, using the data as
received in high time rate mode during the time period 1990 Dec~6 04:29:00
to 04:48:20 (UT).  We added counts in 0.2~s intervals, folded the data
over several likely pulsation periods near the neutron star rotation
period, and applied the Analysis of Variance method (ANOVA, Davies 1990)
to determine the likelihood of pulsations.  No significant pulsations were
found in the lines or continuum.  However, due to the low count rate and
the extensive airglow contamination, the limits on pulsation amplitude, 15
-- 35\% of the total flux, are not stringent. 
 
\section{Discussion}
 
\subsection{Comparison with Previous Observations}
 
The detection of O\,VI in Her X-1 is important, but with a single
observation at a particular $\phi_{\rm orb}$ and $\phi_{35}$,
we are limited in the conclusions we can draw about the system.
In this section we discuss the HUT observation in the context of the
more extensive IUE and HST spectroscopy of Her X-1.
 
Howarth \& Wilson (1983a,b; hereafter HWa and HWb) discuss a consistently
reduced set of IUE data on Her X-1 that cover a range in both orbital
period and the 35 day X-ray cycle.  The line and continuum strengths show
a clear dependence on orbital phase (Figure~3 of HWa displays the
1500~\AA\ continuum and Figure~2 of HWb displays the N\,V and C\,IV line
strengths). Her~X-1 has shown similar line strengths at corresponding
orbital phases in subsequent HST observations (Anderson et al. 1994,
Boroson et al. 1996) and IUE observations (Vrtilek \&\ Cheng 1996).
Plotting the HUT data on Figure~3 of HWa, Figure~2 of HWb, and Figure~5 of
Vrtilek \&\ Cheng demonstrates that the HUT fluxes are about a factor of~2
lower than those observed with IUE at phase 0.34/0.66 (i.e. 0.34 from
mid-eclipse).  Inspection of optical magnitudes provided by the American
Association of Variable Star Observers (AAVSO) shows that the 1.7~day
optical modulation continued during the period surrounding our observation
(V magnitudes $V=13.1$ at $\phi_{\rm orb}=0.40$ and $0.61$, and $V=14.1$
at $\phi=0.98$).  As X-ray heating is thought to cause both the optical
modulation and the UV emission, this suggests that the decrease in UV flux
that we observed was a short-lived phenomenon or unrelated to X-ray
heating.  From our examination of the data we are confident that the 
decrease is not due to any known shortcoming of the data acquisition and 
reduction.  

\subsection{Analysis of the UV Continuum}
 
In addition to the line fits described in \S2, we have produced fits to
the continuum after airglow subtraction.  The continuum spectrum below
$\approx1200$\AA, reported here for the first time, is important because
the accretion disk should contribute more to the continuum flux at shorter
wavelengths. 

In fitting the continuum, we used regions free of strong airglow
emission lines. We chose to fit the continuum with 1) a power law plus
extinction component, and 2) a model of an X-ray illuminated accretion
disk and Roche lobe-filling star. These fits will be described briefly
below. 
 
For the power law fit, we specified initial guesses at a normalization,
$A$ (the flux at 1000~\AA), and the exponent of the power law, $\alpha$,
so that $F=A(\lambda/1000)^{-\alpha}$.  We also applied a Seaton (1979)
galactic extinction law, extropolated to the Lyman limit in a manner that
is consistent with Voyager measurements (Longo et al. 1989).  For the Her
X-1 HUT spectrum, we find $A = 1.3\pm 0.3 \times 10^{-13}$ erg s$^{-1}$
cm$^{-2}$ \AA$^{-1}$, $\alpha = 1.33 \pm 0.17$, and $E(B-V) = 0.09 \pm
0.02$.  The reduced $\chi^2$ for this fit was $1.17$ (with 1125 points and
three free parameters).  The $E(B-V)$ value is inconsistent with UV
observations of the 2200~\AA\ feature (Gursky et al. 1980).  Setting
$E(B-V)=0.05$, we find $A=8.3\pm0.2\times10^{-14}$ and
$\alpha=0.99\pm0.06$.  This fit is nearly as good (the reduced $\chi^2$ is
still $1.17$) as the fit with $E(B-V)$ as a free parameter.  We plot this
latter model against the data in Figure~5. 
 
We also tried fitting the HUT spectrum with a detailed physical model of
an X-ray illuminated accretion disk and star (assumed to fill its Roche
lobe).  The model is identical to that presented by Cheng, Vrtilek, \&
Raymond (1995), except that we have used Kurucz model atmospheres (Kurucz
1992) instead of our library of stellar spectra, which does not extend to
the Lyman limit. The model solves for the temperature structure of the
disk using the method described by Vrtilek et al. (1990), and includes
X-ray heating of the normal star, the eclipse of the star by the disk, and
the X-ray shadow of the disk on the surface of the normal star using the
methods described in HWa.  We used values for the fixed parameters
describing Her~X-1 given in Table~2 of Cheng et al. (1995), including
$E(B-V)=0.0$.

In Figure~5 we compare the continuum spectrum produced by the detailed disk 
model with the power-law fit and with the spectrum observed with HUT.
The disk contributes $\sim10$\%\ of the optical flux, $\sim15$\%\ of the 
flux near 1200\AA, $\sim25$\%\ of the flux near 1100\AA, and
$\sim50$\%\ of the flux near 950\AA.

The only free parameter in the model is $\dot{M}$, the mass transfer rate
onto the neutron star.  The best-fit value for $\dot{M}$ is
$\log_{10}\dot{M}=-8.74\pm0.01$ (in units of solar masses per year). This
value of $\dot{M}$ is lower than the values reported by Vrtilek \&\
Cheng (1996) during an ``anomalous low'' state of Her~X-1 (in their 
final fit, the lowest value is $\log \dot{M}=-8.68$).
With $\dot{M}$ fixed at the
best-fit value, the model predicts $V=13.6$ also at $\phi=0.41,0.61$, when
AAVSO observers report $V=13.1$. 

The fit of the detailed model is worse than the simple power-law; we find
a reduced $\chi^2$ of 1.64.  The fit is especially poor below 1200\AA. We
investigated several possible systematic errors that could have caused
this poor fit: the value we have used for the interstellar reddening, the
flux calibration of HUT, the timing of the 35-day cycle, and the
subtraction of the background airglow spectrum. 

If $0<E_{B-V}<0.05$, then a greater value of $\dot{M}$ is required to fit the
data. The higher resulting temperature of the heated disk and star can
compensate for the spectral change introduced by the reddening.  For
$E_{B-V}=0.05$, the model predicts an observed flux from 1000--1100\AA\
that is $\approx15$\%\ greater than in the case with no reddening;
however, the fit is still not good, and $\chi^2=1.59$. 

Increasing the flux in the HUT spectrum by a factor of 2 to reflect some
unknown flux calibration problem results in a better fit ($\chi^2=1.33$)
and $\log \dot{M}=-8.48$, which is more typical of Her~X-1. At
$\phi=0.41,0.61$, $V=13.4$.  The spectrum predicted by this model is
shown in Figure~5. 

The 35-day cycle of Her~X-1 does not repeat exactly (Baykal et al. 1993).
Thus there is some uncertainty in our choice of $\phi_{35}$, which
determines the orientation of the accretion disk.  If we allow $\phi_{35}$
to vary as a free parameter, we find $\chi^2=1.28$ for the best-fit values
$\phi_{35}=0.15\pm0.01$, $\log \dot{M}=-8.85\pm0.01$. Although the fit is 
much improved, $\log \dot{M}$ is further out of the range of previous 
observations. 
Using this 35-day phase, more of the hot, central disk is visible,
and so the flux at shorter wavelengths is increased.  The uncertainty
in extrapolating the 35-day ephemeris through $\Delta N$ cycles
can be approximated as 0.9 $\Delta N^{1/2}$~d (Baykal
et al. 1993).  The difference between the best-fit value of $\phi_{35}$
and that expected from the nearest reported X-ray turn-on is $\approx8$d,
or a deviation of $\approx3\sigma$.

As described in \S2, we subtracted from the Her~X-1 spectrum a smooth
model of the airglow spectrum based on a blank-sky observation.  There is
marginal evidence in the blank-sky spectrum for weak airglow lines of
Ly$\delta,\epsilon,\gamma$ near 930,938,950\AA, where our fit to the data
is poor.  We tested our method of background subtraction by smoothing the
blank-sky spectrum with a Savitzsky-Golay filter.  While this method
introduces errors near the wings of the strong airglow lines, it can
approximate the airglow background even when strong lines are not present. 
Using this method of background subtraction does not change our derived
value of $\dot{M}$, and the model still predicts too low a flux below
1200\AA, but now $\chi^2=1.39$.  Furthermore, this shows that there may be
significant errors introduced by the background-subtraction.

In summary, we find that the Her~X-1 far-UV spectrum was unusually faint
at the time of our observation, resulting in model fits with values of
$\dot{M}$ lower than previously reported.  The flux $<1200$\AA\ is 
stronger than expected from detailed models.  Uncertainties in the 35-day 
phase, the centering of Her~X-1 in the aperture, or the 
background-subtraction may account for these discrepancies.

\subsection{Analysis of the Emission Lines}

We have compared the observed emission line fluxes with those predicted by
theoretical models of X-ray illuminated accretion disks (Raymond 1993).  
The results are shown in
Table~1.  The ``COS'' model for the emission lines uses cosmic abundances,
while the ``CNO'' model uses a fixed enhancement of He and N and a fixed 
depletion of C to 
represent abundances due to CNO processing.  In computing the emission line
strengths expected from the models, we have assumed that Her~X-1 is at a
distance of 5.5~kpc, and have made no correction for interstellar
reddening.  

The spectrum confirms the basic predictions of the X-ray 
illuminated accretion models.  In particular, the interpretation of the 
N\,V/C\,IV ratio as an element abundance effect, rather than a 
problem with the predicted ionization or density structure, is borne
out by the line strengths.
While the \NV\ doublet is stronger than \CIV, the
\OVI\ lines may be slightly weaker.  This supports the conclusion that
the unexpected strength of the N\,V lines relative to C\,IV does not result 
from a dependence of line flux on ionization potential.  

The presence of O\,VI emission lines in Her X-1 is interesting because
O\,VI is formed in a hotter region than any of the other UV lines.
The ratio of \OVI\ to O\,V$\lambda1371$ provides a temperature diagnostic.
The O\,V line is produced from recombination of O\,VI and thus 
has the same dependence on the O\,VI ionization fraction and on the
electron density as the \OVI\ lines.  The
strength of the \OVI\ lines depends on the strongly temperature-dependent
excitation rate of the upper level of the transition.  Using the
O\,VI recombination rates of Nussbaumer \&\ Storey (1984) and the
standard excitation rate expressions and cross-section found in Osterbrock 
(1989), we find temperatures of $1.44\pm0.07\times10^{4}$~K
and $1.48\pm0.07\times10^{4}$~K for the \OVI\ fluxes inferred
assuming, respectively, a 2:1 doublet ratio and a 1:1 ratio with 
interstellar absorption.  Correcting for reddening (using $E_{B-V}=0.05$)
increases O\,VI/O\,V by a factor of 1.44.  This only increases the 
temperature estimate to 15,000~K.

HWb model the line emitting region, and for the N\,V region find $\log T 
= 4.44 \pm  0.02$ , and $\log n  = 13.24 \pm  0.06$,
assuming a cosmic abundance of N and that all N is in the N\,V  state.
Our temperature estimate has assumed that the \OVI\ lines are
optically thin; if this assumption is not right, then the temperature may
be much higher.  Further, we have assumed that the O\,VI lines and the
O\,V line are emitted in the same regions or in regions with similar
temperatures and densities, which may not be the case.

Both the COS model and the CNO model predict observable  
C\,III$\lambda977$ emission (7.5$\times10^{-13}$ and 4.2$\times10^{-13}$ 
erg~s$^{-1}$~cm$^{-2}$ respectively).  The weak features that we 
saw had total flux $<2.5\times10^{-13}$ erg~s$^{-1}$~cm$^{-2}$.
Interstellar C\,III$\lambda977$ absorption would probably decrease
the strength of this line by $\lae20$\%.
We note that Anderson et al. (1994) observed the C\,III$\lambda1175$ 
line with HST (when the N\,V and C\,IV lines were twice as strong as
observed with HUT), and found a flux of 
$2.7\times10^{-13}$~erg~s$^{-1}$~cm$^{-2}$, whereas COS and CNO predict
fluxes of $1.0\times10^{-12}$ and $4.6\times10^{-13}$~erg~s$^{-1}$~cm$^{-2}$,
respectively.  A further decrease in the C/N ratio beyond $C/N=0.66$ as 
assumed in model CNO to C/N$=0.5$ would improve the agreement between 
observations and models, but the evidence is not overwhelming.

This observation with the Hopkins Ultraviolet Telescope was the first 
to study the flux from this interesting source below the Lyman~$\alpha$ line,
where a large fraction the continuum emission probably arises in the disk, 
and where there are strong emission lines especially sensitive to X-ray
illumination.  Future observations with FUSE will be able to 
detect these features with greater sensitivity and resolution.
 
\acknowledgements
 
It is a pleasure to thank the many members of the HUT team at JHU and at
the JHU Applied Physics Laboratory whose years of work paid off in two
successful Astro missions.  We also thank the Spacelab Operations Support
group at Marshall Space Flight Center and the crew of the Astro-1 mission
for their support before and during the flight.  We would like to thank
the referee, Ian Howarth, for a careful reading and for suggestions on
improving the manuscript.  The Hopkins Ultraviolet Telescope project is
supported by NASA contract 5-27000 to The Johns Hopkins University.  BB
and SV were supported in part by NASA (NAG5-2532, NAGW-2685), and NSF 
(DGE-9350074).

In this research, we have used, and acknowledge with
thanks, data from the AAVSO International Database,
based on observations submitted to the AAVSO by variable
star observers worldwide.

\clearpage
 
\begin{table}
\caption{HUT Measurements of Emission Lines in
Her X-1} 
\vspace{0.5in}
\scriptsize
\begin{tabular}{lcccrcccccc} 
Line & $\lambda_{\rm obs}$\tablenotemark{a}& $\lambda_{\rm lab}$ & 
FWHM($\lambda$)
& Blueshift & Flux & COS\tablenotemark{b} & CNO\tablenotemark{c} 
& $\chi^2_{\rm red}$\\
 & (\AA) & (\AA) & (\AA) & (km/s) & (erg s$^{-1}$ cm$^{-2}$) & & &\\ 
\tableline
O\,VI & 1029.28$\pm0.19$ & 1031.93 & $3.60\pm0.36$ & 
770$\pm60$ &
	7.1$\pm0.7(-13)$ & 1.8(-12) & 8.9(-13) & 78.3/44\\ 
O\,VI\tablenotemark{d} & 1029.12$\pm0.14$ & 1031.93 & $3.61\pm0.35$ & 
820$\pm40$ &
	8.9$\pm0.1(-13)$ & & & 66.0/42 \\
N\,V & 1236.95$\pm0.45$ & 1238.82 & 8.7$\pm2.6$ & 450$\pm110$ & 
8.3$\pm1.0(-13)$ & 4.7(-13) & 5.9(-13) 
        & 29.6/35\\
C\,IV & 1547.16$\pm0.58$ & 1548.20 & 7.0$\pm1.4$ & 200$\pm110$ &
3.27$\pm0.51(-13)$ & 1.03(-12) & 5.3(-13) & 94.1/94\\
O\,V & 1369.23$\pm0.18$ & 1371.29 & $1.4\pm0.4$ & 450$\pm40$ &
	8.2$\pm2.4(-14)$ & 6.1(-14) & 4.4(-14) & 40.1/33 \\
C\,III &  & 977.02 & & & $<2.5$(-13) & 7.5(-13) & 4.2(-13) & \\
\end{tabular} 
\normalsize 
\tablenotetext{a}{For doublets, the wavelength given is the wavelength
of the blue component.}
\tablenotetext{b}{Model with Cosmic abundances, from Raymond (1993), 
Table~2.  The distance to Her~X-1 is assumed to be 5.5~kpc.}
\tablenotetext{c}{Model with metal abundances enhanced due to CNO
processing. From Raymond (1993).}
\tablenotetext{d}{Assuming an interstellar absorption line and
an intrinsic 1:1 doublet ratio.  The flux listed is the flux in the
unabsorbed line.}

\end{table}

\clearpage
 
\figcaption{ 
Flux-calibrated HUT spectrum of Her X-1.  The top curve shows the observed 
spectrum before airglow subtraction, smoothed over five pixels (= 2.5\AA)
for display purposes.  
The bottom curve shows a smooth fit to a pure airglow spectrum 
obtained with the same slit but at a different time during the mission.}  
 
\figcaption{
Observed HUT spectrum of Her X-1 after first-order removal of airglow. 
This spectrum has been smoothed over five pixels.}
  
\figcaption{The observed HUT spectrum in the vicinity of \OVI, the locally 
scaled background airglow emission, and our fit to the airglow lines (bold).
}

\figcaption{
Plots of Her X-1 emission lines showing fits to the doublet profiles as 
indicated in Table 1.  a) O\,VI region; b) N\,V region; c) C\,IV region;
and d) O\,V region.
The data are shown unsmoothed.  The vertical lines represent the rest
wavelengths of the emission lines.}
  
\figcaption{The UV continuum flux observed by HUT (smoothed over 7 
pixels) compared with a power-law 
fit (dotted line, labelled ``PL''), a model of an X-ray illuminated star and 
disk (solid line, labelled ``Model'').  We also 
show a model fit to twice the observed spectrum and then divided by 2 
(dashed line) to show the effect of a possible underestimate of the true 
flux from Her~X-1.}

\newpage

\center{\bf Figure~1}
\vspace{3mm}
\begin{figure}[h]
\plotone{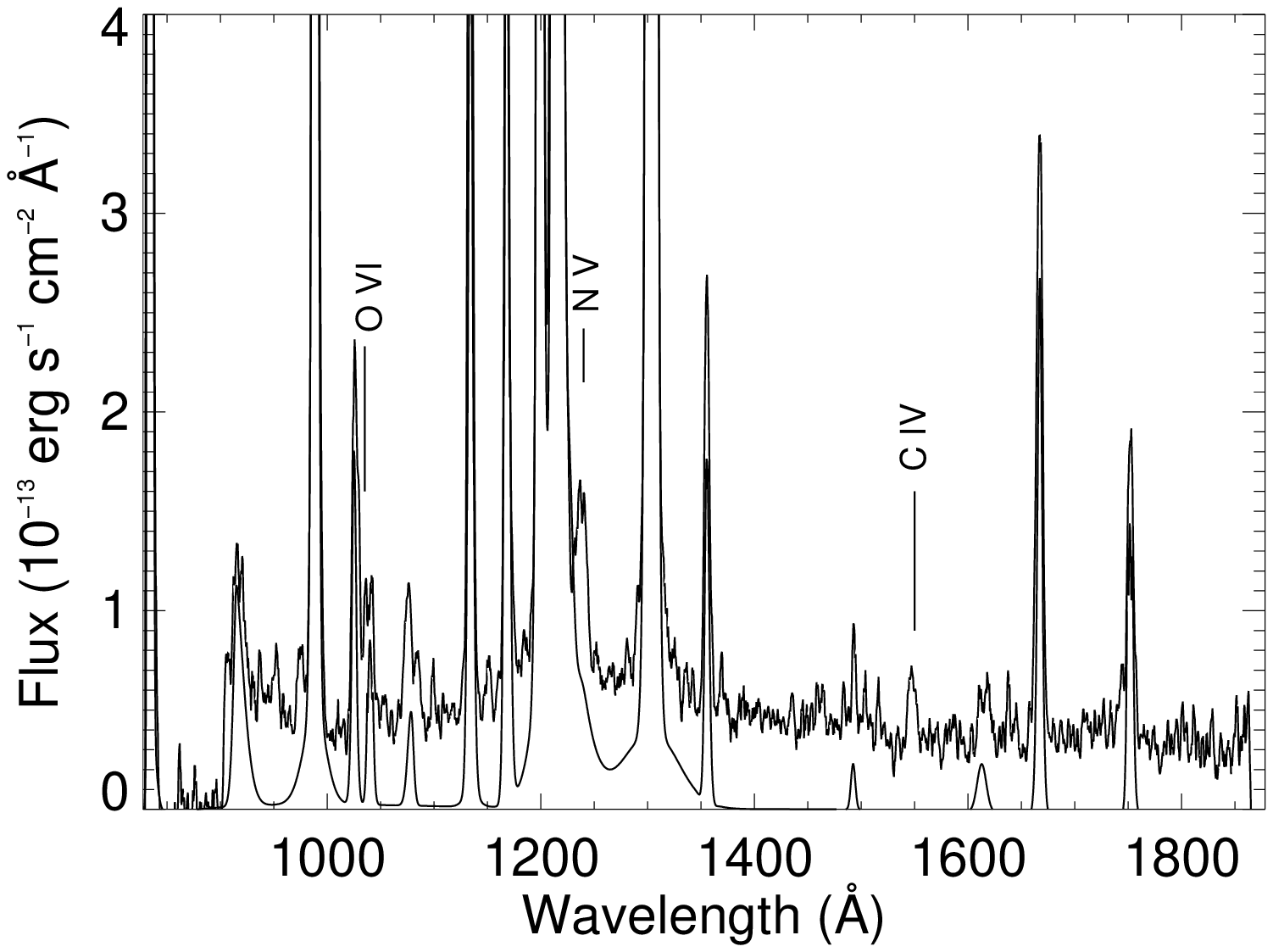}
\end{figure}
\newpage

\center{\bf Figure 2}
\vspace{3mm}
\begin{figure}[h]
\plotone{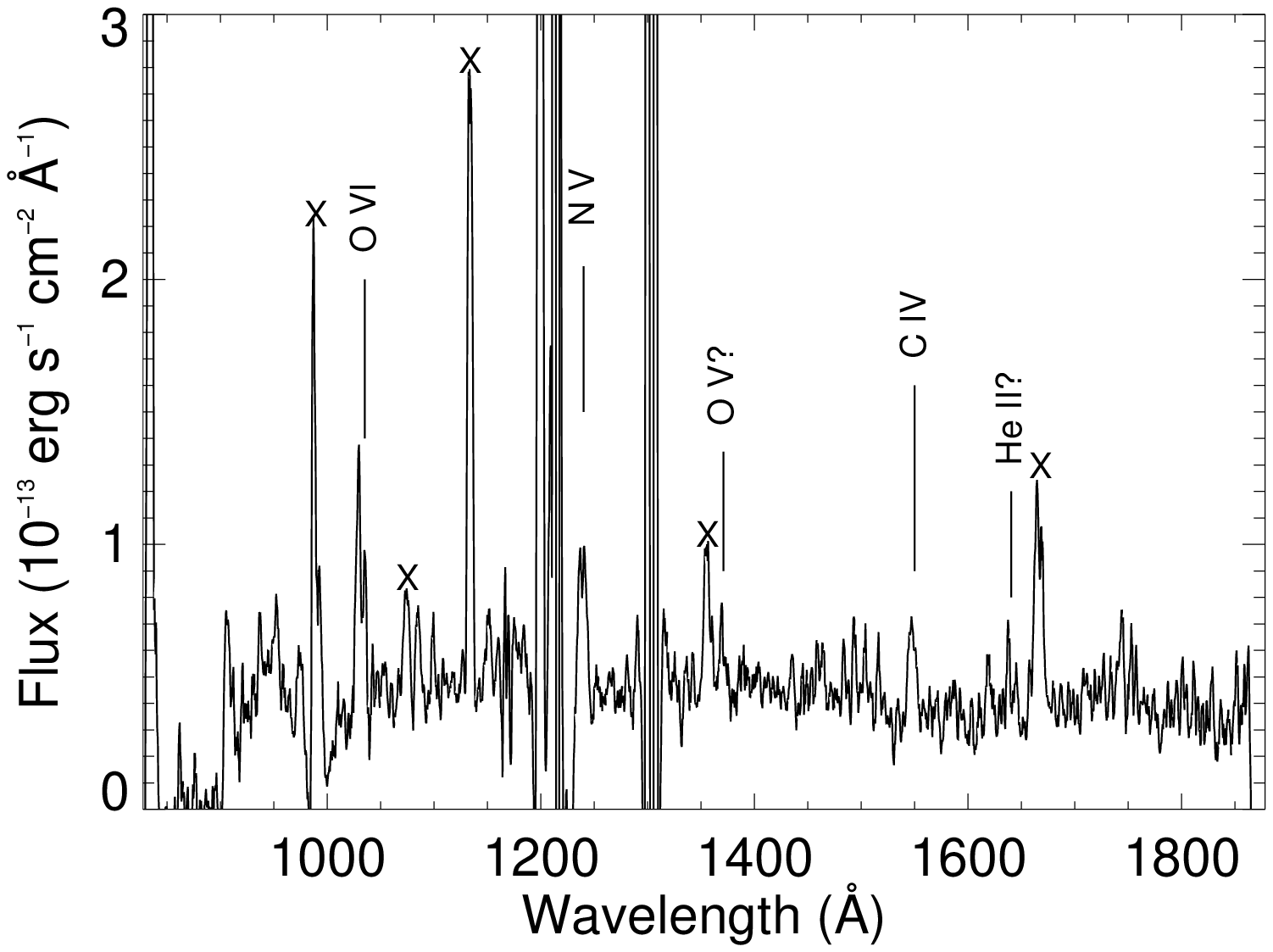}
\end{figure}
\newpage

\center{\bf Figure 3}
\vspace{3mm}
\begin{figure}[h]
\plotone{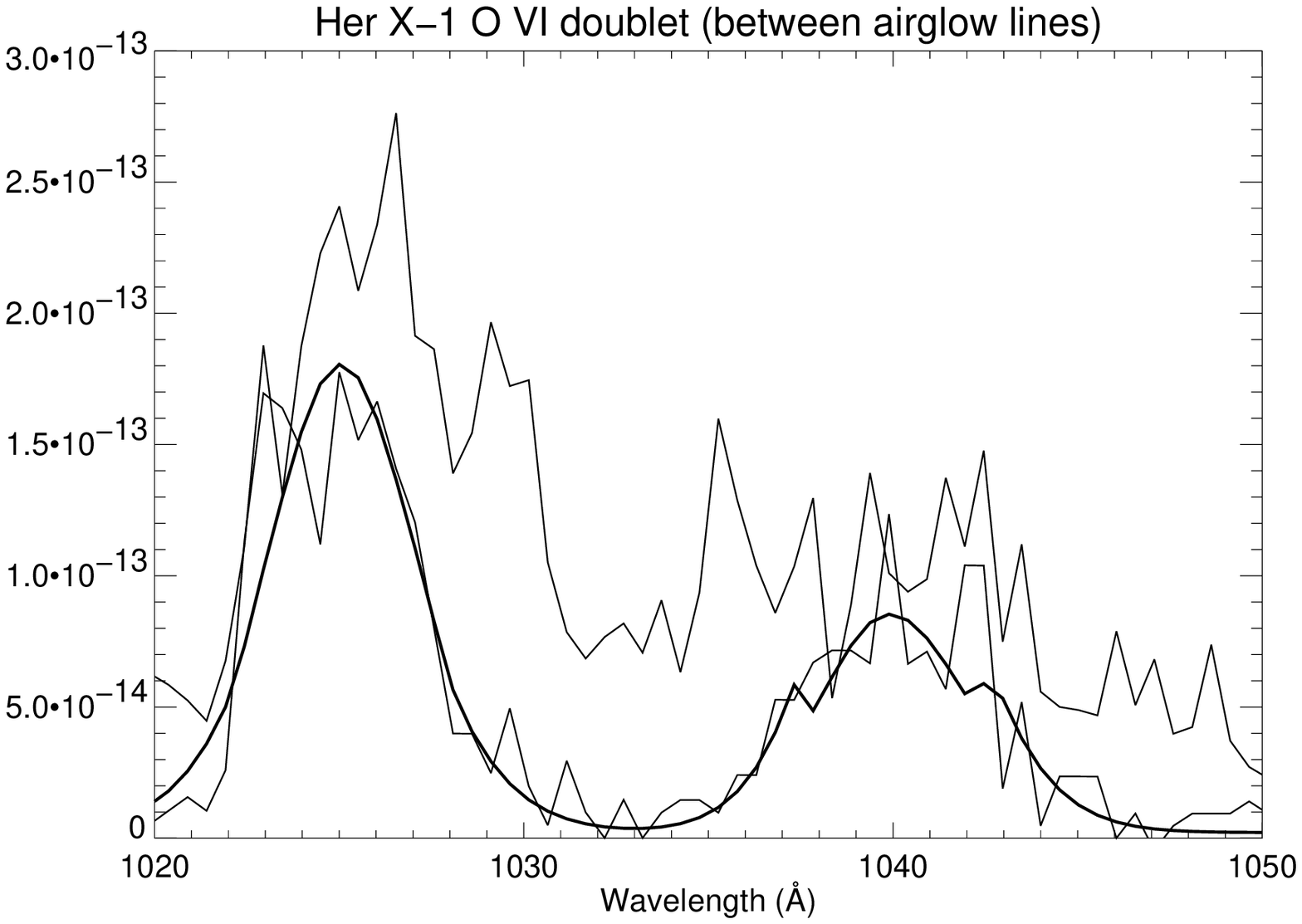}
\end{figure}
\newpage

\center{\bf Figure 4}
\vspace{3mm}
\begin{figure}[h]
\plotone{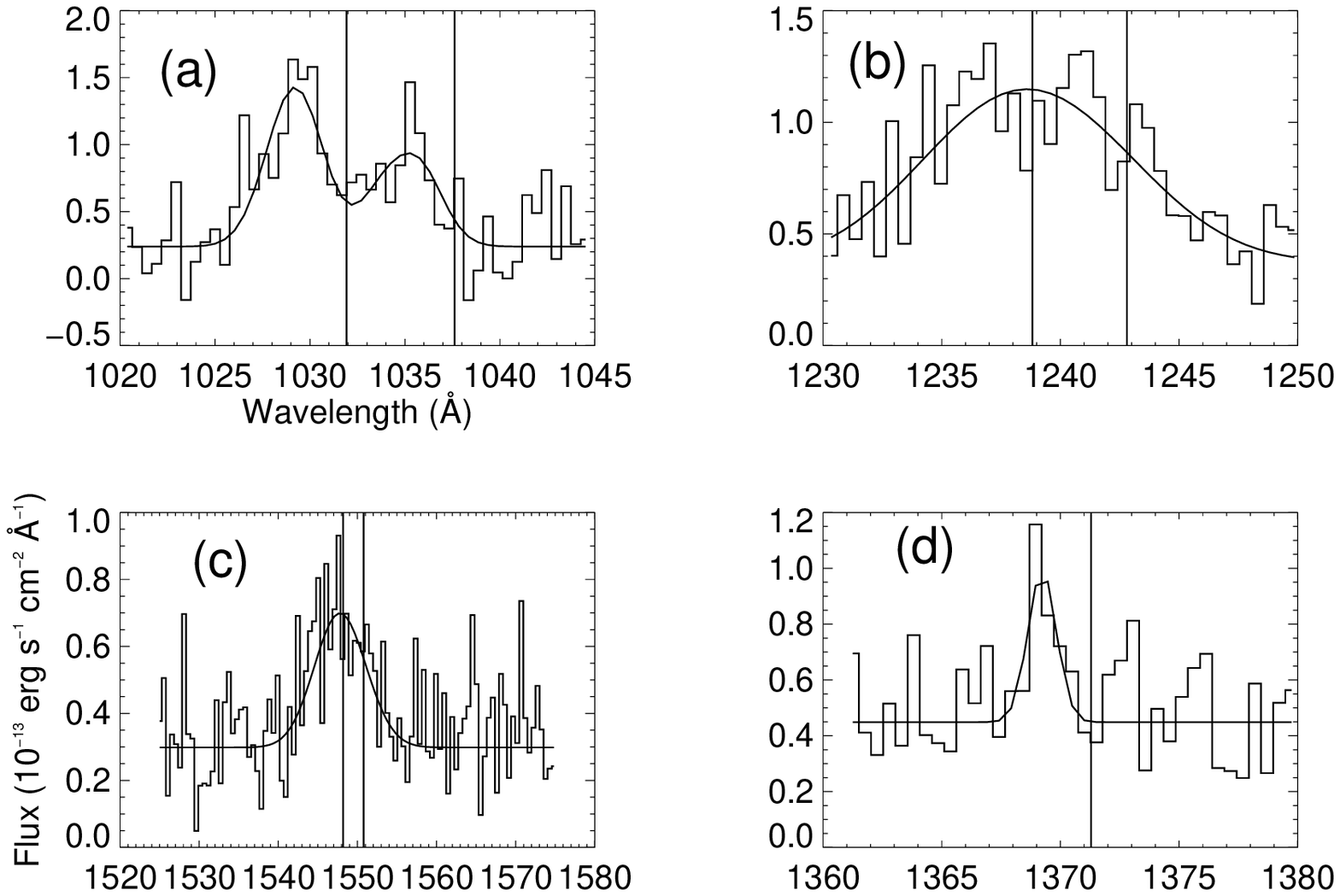}
\end{figure}
\newpage

\center{\bf Figure 5}
\vspace{3mm}
\begin{figure}[h]
\plotone{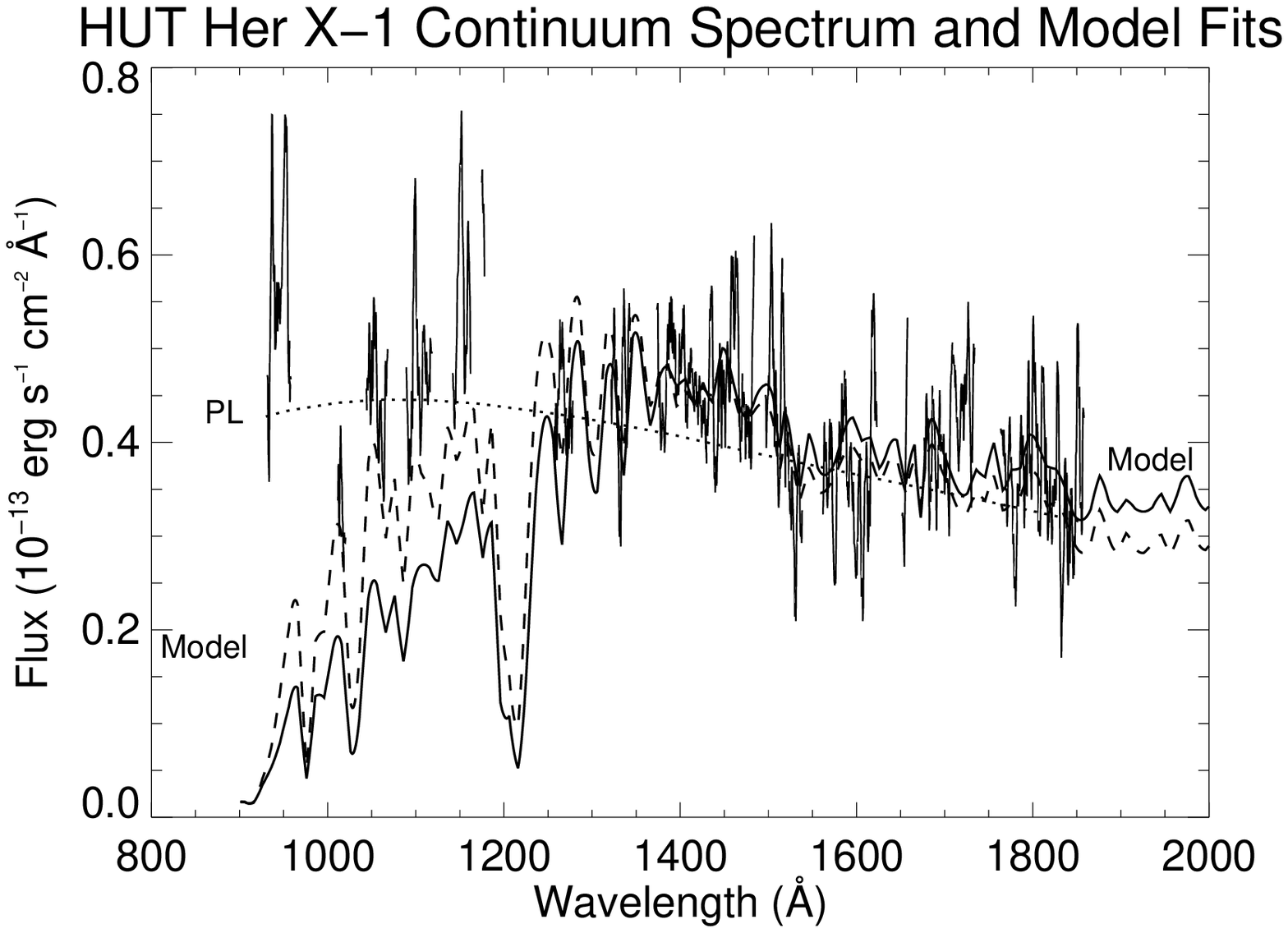}
\end{figure}


\clearpage 
  
\begin{references}

Anderson, S.F., Wachter, S., Margon, B., Downes, R.A. 1994, ApJ, 
436, 319

Bahcall, J.N., \& Bahcall, N.A.  1972, ApJ, 178, L1

Baykal, A., Boynton, P.E., Deeter, J.E., \& Scott, D.M.  1993, MNRAS 
265, 347

Blair, W. P., \& Gull, T. R. 1990, S\&T, 79, 591

Boroson, B., Vrtilek, S.D., McCray, R., Kallman, T., \& Nagase, F.
1996, ApJ, 473, 1079
 
Boyle, S., Howarth, I., Wilson, R., \& Raymond, J.  1986, in ESA 
Proc. Int. Symp. on New Insights in Astrophysics, 471
 
Cheng, F.H., Vrtilek, S.D., \& Raymond, J.C.  1995, ApJ, 452, 825
 
Davidsen, A. F., Henry, J. P., Middleditch, J., \& Smith, H. E. 
1972, ApJ, 177, L97
 
Davidsen, A. F. et al. 1992, ApJ, 392, 264
 
Davies, S.R. 1990, MNRAS, 244, 93
 
Deeter, J. E., Boynton, P. E., Miyamoto, S., Kitamoto, S., 
Nagase, F., \& Kawai, N. 1991, ApJ, 383, 324
 
Dupree, A. K., et al. 1978, Nature, 275, 400
 
Feldman, P. F., et al. 1992, Geophys. Res. Let., 19, 453
 
Gerend, D., \&\ Boynton, P. E. 1976, ApJ, 209, 562
 
Giacconi, R., Gursky, H., Kellogg, E., Levinson, R., Schreier, 
E., \& Tananbaum, H. 1973, ApJ, 184, 227
 
Gruber, D. E., et al. 1980, ApJ, 240, L127
 
Gursky, H., et al. 1980, ApJ, 237, 163
 
Howarth, I. D., \&\ Wilson, B. 1983a, MNRAS, 202, 347 (HWa)
 
Howarth, I. D., \&\ Wilson, B. 1983b, MNRAS, 204, 1091 (HWb)
 
Kondo, Y., Van Flandern, T. C., \&\ Wolff, C. L. 1983, ApJ, 
273, 716

Kruk, J.W., Kimble, R.A., Buss, R.H., Davidsen, A.F., Durrance, S.T.,
Finley, D.S., Holberg, J.B., and Kriss, G.A.  1997, ApJ, 482, 546

Kurucz, R.L.  1992, in The Stellar Populations of Galaxies, ed. B Barbuy 
\&\ A. Renzini (Dordrecht: Klower), p. 225

Longo, R., Stalio, R., Polidan, R. S., \&\ Rossi, L. 1989, ApJ, 
339, 478
 
Meyer, F., \& Meyer-Hofmeister, E.  1984, A\&A 104, 35
 
Middleditch, J. 1983, ApJ, 275, 278
 
Middleditch, J., Puetter, R. C., \& Pennypacker, C. R. 1985, ApJ, 
292, 267
 
Milgrom, M., \&\ Salpeter, E. E. 1975, ApJ, 196, 589

Nussbaumer, H., \& Storey, P.J.  1984, A\&AS, 56, 293
 
\"Ogelman, H. 1987, A\&A, 172, 79

Osterbrock, D.E.  1989, Astrophysics of Gaseous Nebulae and 
Active Galactic Nuclei (Mill Valley,, California: University Science 
Books)

Petterson, J.A.  1975, ApJ, 201, L39

Raymond, J.C.  1993, ApJ, 412, 267

Schandl, S., \& Meyer, F.  1994, A\&A 289, 149
 
Seaton, M. J. 1979, MNRAS, 187, 73P
 
Tananbaum, H. Gursky, H., Kellogg, E. M., Levinson, R., Schreier, 
E., \& Giacconi, R. 1972, ApJ, 174, L143
 
Voges, W., Pietsch, W., Reppin, C., Trumper, J., Kendziorra, 
E., \& Staubert, R., 1982, ApJ, 263, 803

Vrtilek, S.D., \&\ Cheng, F.H.  1996, ApJ, 465, 915
 
 
Wolff, C. L., \&\ Kondo, Y.  1978, ApJ, 219, 605
 
\end{references}
\end{document}